\documentclass[9pt,twocolumn,twoside]{article}
\usepackage[utf8]{inputenc}
\usepackage{authblk}
\usepackage{float}
\usepackage{siunitx} 
\usepackage{graphicx}
\usepackage[margin={1in,1in}]{geometry}

\title{Harvesting Planck radiation for free-space optical communications in the LWIR band}

\author[1,2,*]{Haley A. Weinstein}
\author[1]{Zhi Cai}
\author[1,2]{Stephen B. Cronin}
\author[1,2]{Jonathan L. Habif}

\affil[1]{Ming Hsieh Dept. of Electrical \& Computer Engineering, University of Southern California, Los Angeles, California 90089, USA}
\affil[2]{Information Sciences Institute, University of Southern California, Waltham, Massachusetts 02451, USA}

\affil[*]{Corresponding author: haweinst@usc.edu}

\begin{document}

\maketitle

\begin{abstract}
We demonstrate a free-space optical communication link with an optical transmitter that harvests naturally occurring Planck radiation from a warm body and modulates the emitted intensity. The transmitter exploits an electro-thermo-optic effect in a multi-layer graphene device that electrically controls the surface emissivity of the device resulting in control of the intensity of the emitted Planck radiation. We design an amplitude-modulated optical communication scheme and provide a link budget for communications data rate and range based on our experimental electro-optic characterization of the transmitter. Finally, we present an experimental demonstration achieving error-free communications at 100 bits per second over laboratory scales.
\end{abstract}

\section{Introduction}
Free-space optics (FSO) is an important modality for communications in scenarios requiring \emph{ad hoc} channels able to operate in spectral bands largely unmanaged by regulatory agencies and have properties that can achieve low-probability of detection (i.e., covert communication). Laser-based FSO communications is argued to have improved power efficiency and drastically improved stealth over RF modalities, owing to the ability to achieve narrow collimation of laser beams improving optical transmitter antenna gain, which improves the coupling efficiency between the transmitter and receiver, and concentrates the optical energy within the atmospheric channel between the transmitter and receiver \cite{Ricklin}. This approach to efficiency and stealth, however, requires precise pointing acquisition and tracking at the transmitter, which often precludes a transmitter operating in a small size, weight and power form factor.
\begin{figure}
  \centering
  \includegraphics[width=\columnwidth]{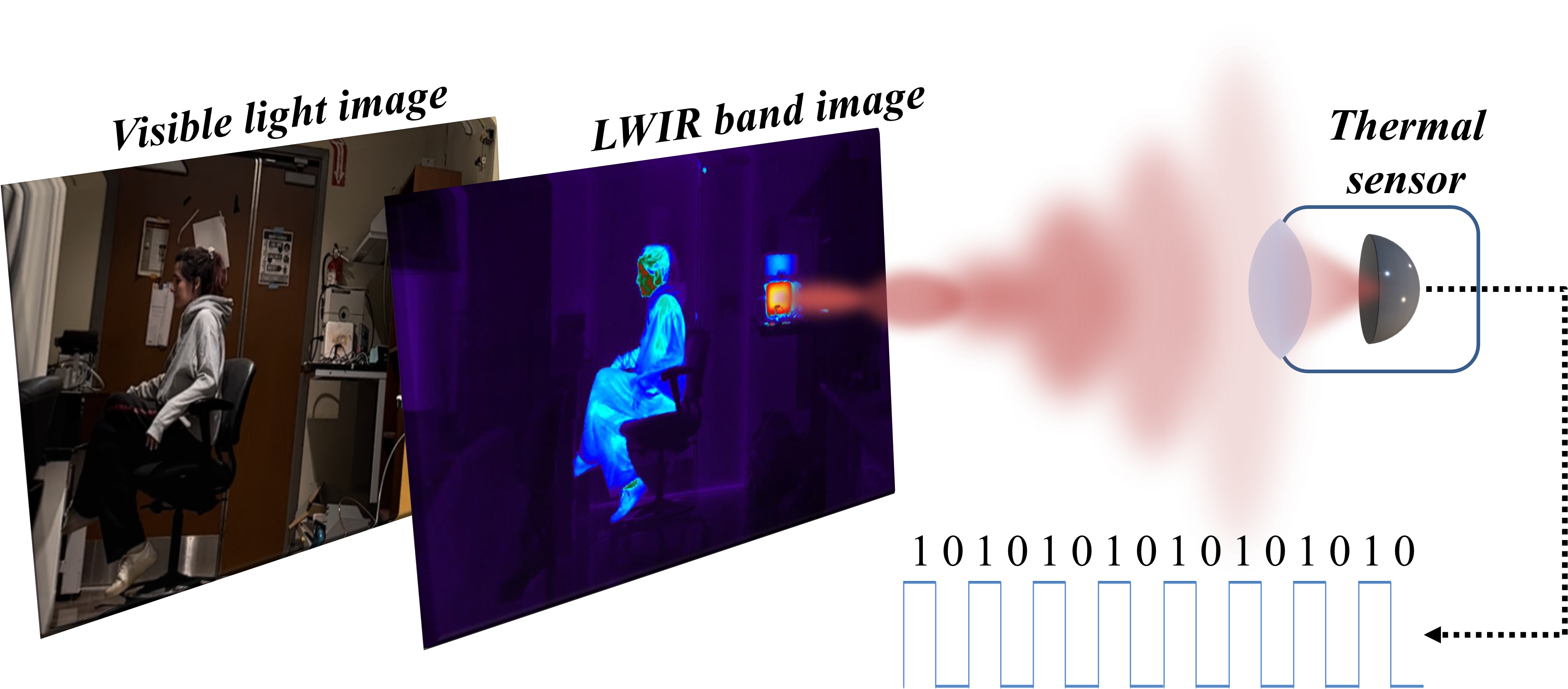}
  \caption{A scene imaged with a thermal camera delivers a spatial map of the Planck radiation emitted by the environment in the field of view. We have demonstrated a communications approach that harvests this Planck radiation and modulates it in time to encode digital data.}
  \label{fig:ConceptDiagram}
\end{figure}
\begin{figure*}
\centering
\includegraphics[width=\textwidth]{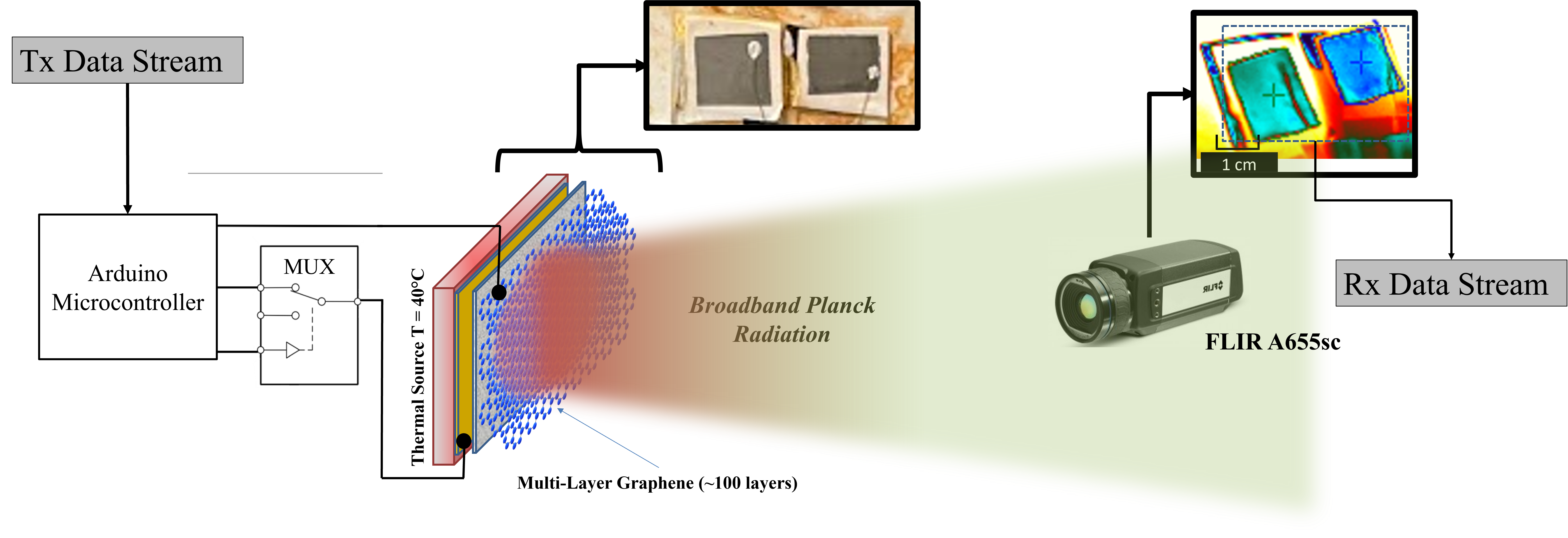}
\caption{Schematic diagram of experimental setup including full diagram and detailed diagram of switching architecture.}
\label{fig:ExptDiagram}
\end{figure*}

An alternative approach to stealthy communications has emerged recently that uses naturally occurring sources of noise to obscure the presence of a communications channel\cite{bash2013limits}, even if the electromagnetic emissions are not contained solely within the optical channel connecting the transmitter and receiver. This approach to stealthy communications has been experimentally demonstrated in the near infrared band \cite{Bash2015} where the authors posit that since the overwhelming abundance of naturally occurring noise in our environment is in the long-wave infrared (LWIR) wavelength band, optical communications in the LWIR would result in the highest performing covert communications link. A channel that operates over the entirety of the LWIR band (7-\SI{14}{\um} ) would provide the greatest flexibility for covert data transmission, contrary to transmitters constructed around a narrowband source, such as a quantum cascade laser \cite{pavelchek2004long} which, in addition, require multiple Watts of power for operation. 

%Following a conventional FSO communications implementation, a quantum cascade laser operating in the LWIR band would be the natural choice for a communications transmitter\cite{pavelchek2004long}. A laser, however, often requires high drive current, and operates in a very narrow spectral band limiting the peak power that can be used while remaining covert, and obviating the value of a laser source. Alternatively, an optical source that radiates its power over the entirety of the LWIR spectral band can operate with higher peak power, as its signature is buried within a richer diversity of environmental noise.6

In this article we demonstrate a FSO communications link with a graphene-based optical transmitter that harvests naturally-occurring Planck radiation from a warm body and modulates the intensity of the Planck emission by electrically modulating the surface emissivity of the graphene device. We characterize the time-dependent electrical modulation of the graphene device surface emissivity using a novel approach to voltage and impedance switching and develop a digital modulation scheme optimized for its electro-optical response. We then demonstrate a FSO communications link implementing a direct detection receiver with a commercially-available vanadium-oxide based thermal imaging camera and demonstrate an error-free communications link over $\sim10$ meters achieving $100$ bps data rate. 

Traditional FSO communications systems are designed using a narrowband, \emph{active} optical source, such as a laser or LED, at the transmitter and a receiver using either direct or coherent detection \cite{Ricklin}. In contrast, our transmitter uses a \emph{passive} approach harvesting Plank radiation from a warm body and modulating its emitted intensity by electrically modulating the surface emissivity of a material layer encapsulating the warm body. Typically, the emissivity of a material is a static parameter depending on wavelength via the value $\epsilon\left(\lambda\right) \in \left[0,1\right]$. In general, however, the emissivity can be a function of a broad host of parameters including temperature \cite{morsy2020experimental}, nano-scale spatial structure \cite{Chanda,morsy2021coupled} and electronic surface potential \cite{Inoue2014,kim2018electronically}. Our communications transmitter design was inspired by electro-chemical engineering of multi-layer graphene (MLG) devices, where the MLG acts as a cathode sandwiching a layer ionic liquid between itself and a conducting cathode \cite{salihoglu2018graphene,Ergoktas2020,zhao2019tunable}. Voltage modulation across the devices enables the rapid electronically controlled switching of the device's surface emissivity over a broad wavelength range covering the LWIR band\cite{Zhi}. These devices have been shown able to modulate the emissivity of a surface enough to change the apparent surface temperature by more than $10$ Kelvin. In this work, we characterize this time-dependent, reversible emissivity modulation to engineer a device for implementation as a communications transmitter. We characterize the electro-thermo-optical response with a room-temperature camera, which we subsequently use as a direct-detection receiver in our communications demonstration. 

\section{Experimental Setup}
\begin{figure*}
  \centering
  \includegraphics[width = 7in]{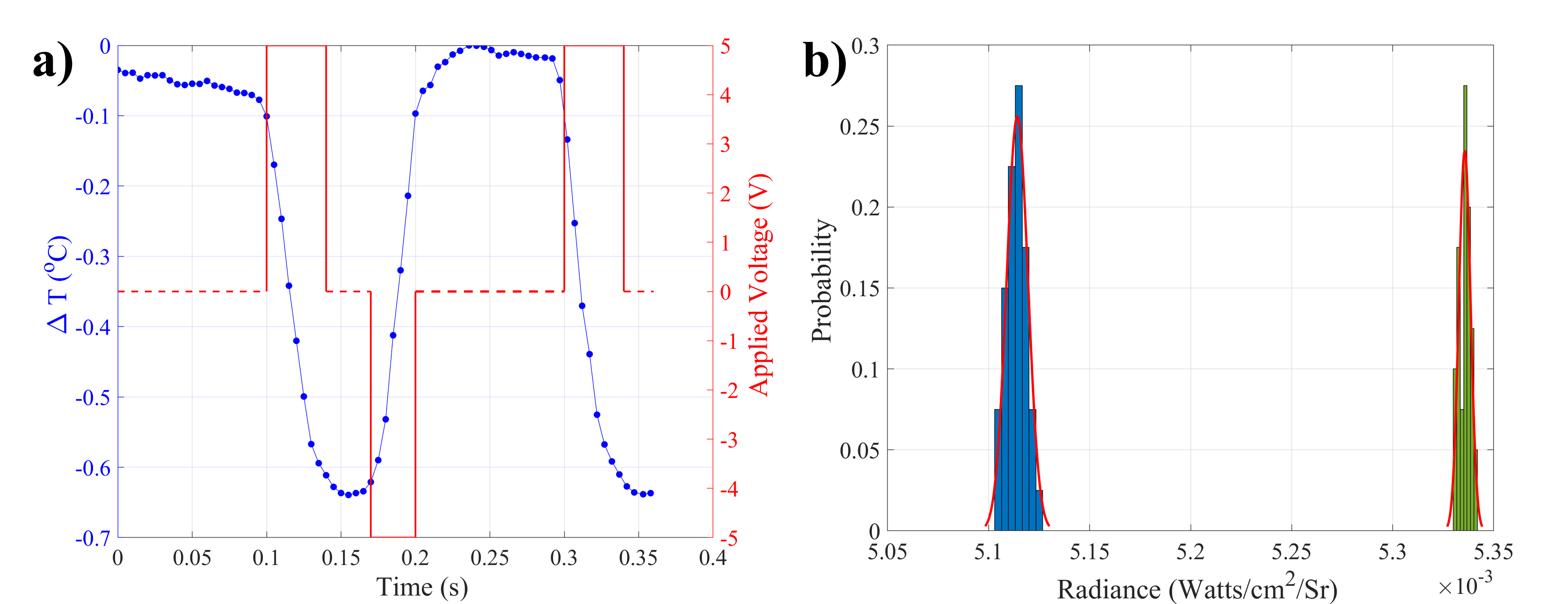}
  \caption{Example characterization of modulated radiance values. \textbf{a)} demonstrates the applied voltage for a 100 ms symbol length OOK signal. A "0" is initially transmitted by switching the MUX to open, this is represented by a dotted line. Then a "1" is transmitted by applying a 30 ms 5 V pulse, dropping the emissivity level, followed by a 20 ms open circuit to hold this level, and returned to nominal emissivity by applying a 30 ms -5 V pulse. \textbf{b)} Photo-statistic characterization curves for transmitting a "1" blue and transmitting a "0" green.}
  \label{fig:Voltage_Signal}
\end{figure*}

The design, fabrication, and material characterization of the Planck modulator devices we have developed follow work from \cite{salihoglu2018graphene} and is described in detail in \cite{Zhi}. The experimental setup for the electro-thermo-optical characterization of the Planck modulator MLG devices is shown in Fig. \ref{fig:ExptDiagram}. The fabricated devices were mounted onto a temperature-controlled hot plate and thermalized to temperatures between $40^{\circ}C$ - $50^{\circ}C$. The Planck modulator device is driven by an Arduino micro-controller and TMUX1248 switch so that we can rapidly change both the applied voltage and the electrical impedance seen by the device. The surface radiance of the device was monitored with a commercially-available thermal imaging camera (FLIR A655sc).  The FLIR camera has sensitivity across the LWIR band ( \SI{7}{\um}-\SI{14}{\um}) and reports internal noise equivalent differential temperature (NEDT) $<30$ mK, though background fluctuations dominated the noise in our measurements. The focal plane array consists of a 640x480 pixel array of un-cooled vanadium oxide microbolometers with a detector pitch of \SI{17}{\um} yielding a field of view (FOV) of $19$\textdegree. The camera has a maximum frame rate of 200 frames per second (fps), which sets the sampling rate for our time-dependent device characterization. The camera software computes the interpreted temperature of the device using the Planck equation\cite{das2015wavelength}, along with user-provided data about the assumed emissivity of the surface being imaged and the distance of the camera from the device. Suppression of the surface emissivity results in suppression of the device radiance and the camera interprets this as a drop in device temperature, even though the ambient temperature of the hot plate remains fixed. When the applied voltage is swept above 3V, we observe an abrupt onset of intercalation of ionic liquid into the MLG. This presents itself as a biexponential time-dependent drop in apparent temperature. 

To characterize the device for use as a communications transmitter, we developed a custom circuit to control the amount of intercalation by applying short voltage pulses of varying time duration to the device followed by an open circuit to freeze the instantaneous emissivity value. Previous works rely on modulating the magnitude of the voltage applied to change the level of intercalation and resultant surface emissivity \cite{Ergoktas2020}. In this work, we found that we can achieve far more stable levels by modulating the time the voltage is applied.

An Arduino microcontroller supplied 5V to the device and controlled a TMUX1248 switch to provide an open circuit when the desired emissivity level was reached to pause the intercalation at this level. This open circuit process stops the movement of anions, setting a stationary emissivity level. While one can drive large emissivity shifts, as shown in \cite{Ergoktas2020}, after applying multi-second voltage pulses, we observed that this intercalation process is not fully reversible and after multiple iterations, the devices can no longer de-intercalate. Figure \ref{fig:Voltage_Signal}a.) shows an example of the voltage applied to the Planck modulator device and the resulting radiance change measured by the camera. At a 5V applied voltage, the maximum level of reversible intercalation occurs at approximately a 40 ms pulse length for our devices. To preserve the longevity and achieve maximum operability, we designed a return-to-zero (RZ) amplitude modulation encoding scheme where, after each positive voltage pulse, we applied a negative voltage pulse of equal duration to the positive pulse to return the device to the initial level of emissivity/intercalation. This ensures the devices will never be pushed to a point where they will no longer be able to de-intercalate. The characterization temperature for the devices was set to 40 \textdegree C, close to body temperature.

In preparation for the communications demonstration, we characterized our ability to resolve radiance values for the modulation of our Planck modulator. Figure \ref{fig:Voltage_Signal}b shows histograms of measurements of the radiance value with the device intercalated and de-intercalated; the high signal to noise ratio (SNR) enables error free communications.

A randomized sequence of 100 bits was generated and the corresponding time-dependent voltage signal was repeatedly applied to the Planck modulator. During operation of the link, the electrical power drawn by the device for switching was approximately $7$ mW, neglecting the power required for the hot plate and the micro-controller.  The FLIR camera uses direct detection to capture the transmitted signal at a frame rate of 200 fps. Image processing functionality within the camera software bins the pixels that are trained on the Planck modulator device and returns an average value of the radiance measured by those pixels. The FLIR camera is controlled using the \emph{ResearchIR} software, which we programmed to acquire a 6 second video. During the video the 100 bit sequence was constantly being transmitted by the device. The video was then post-processed using Matlab, where individual frames of the video were analyzed to extract the transmitted data.

 \begin{figure}
  \centering
  \includegraphics[width=3in]{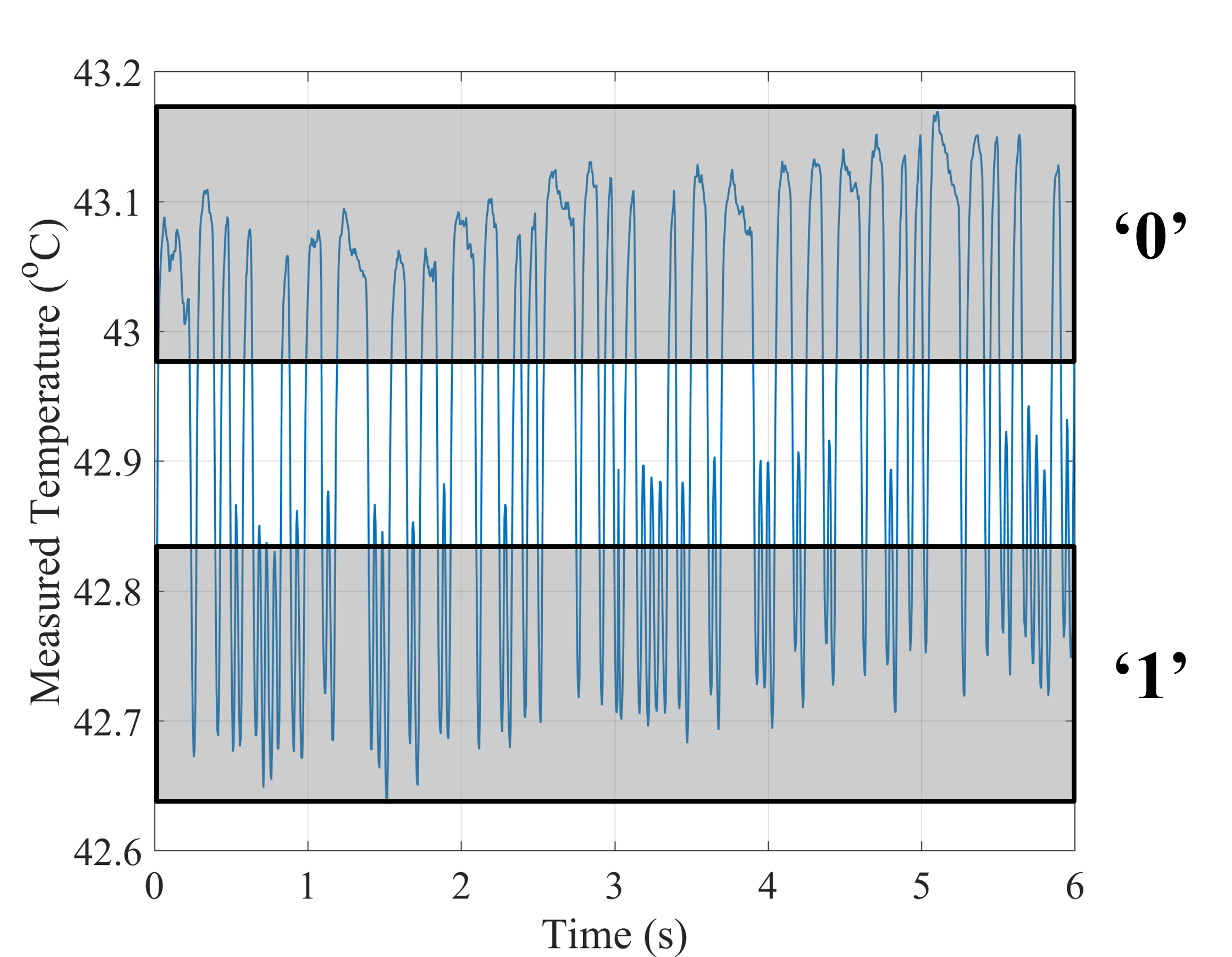}
  \caption{Resulting apparent temperature modulation from On-off-keying (OOK).}
  \label{fig:OOK}
\end{figure}

For our demonstration we enhanced our channel capacity by spatially multi-plexing two Planck modulators, commanded by the same microcontroller, side by side at the transmitter, as illustrated in Fig. \ref{fig:ExptDiagram}. Each Planck modulator had an active area of 1 cm x 1 cm. At a range of 10 meters each Planck modulator is imaged by approximately 1 pixel on the focal plane array of the camera, putting our demonstration just at the edge of the far field. Each device repeatedly transmitted a different 100 bit sequence. 

We began with encoding On-Off-Keying (OOK) data format for transmission. Figure \ref{fig:OOK} shows the measured signal captured by the receiver for one device. This signal and the corresponding signal for the other device were subsequently decoded into a bit string by a MATLAB program which aggregated the amount of time the recorded temperature fell below a dynamic threshold. The threshold was calculated based on the minimum and maximum received temperature, and was recalculated at intervals of 10 bits. This allowed the classification algorithm to be resilient to fluctuations in the temperature of the hot plate and intermittent camera re-calibration. 

To scale our communications data rate towards 100 bps using the OOK data format we decreased the symbol duration from 100 ms (shown in Fig. \ref{fig:Voltage_Signal}a) towards 10 ms. As the optical receiver samples at ~200 fps, we could not reliably decode data that was below a 10 ms symbol length. We found that we could demonstrate error free communication from 1 to 10 meters at symbol lengths $>$20 ms. With a higher frame rate camera this symbol length could decrease further yielding higher data rates. 

\section{Discussion and Results}

Using 20 ms symbol length OOK with spatial multiplexing of two devices, we repeatedly transmitted a random sequence of 100 bits. This corresponded to a rate of 100 symbols per second. Using the aforementioned MATLAB script, we decoded the sequence achieving error free communications. 

\begin{figure}
  \centering
  \includegraphics[width=3in]{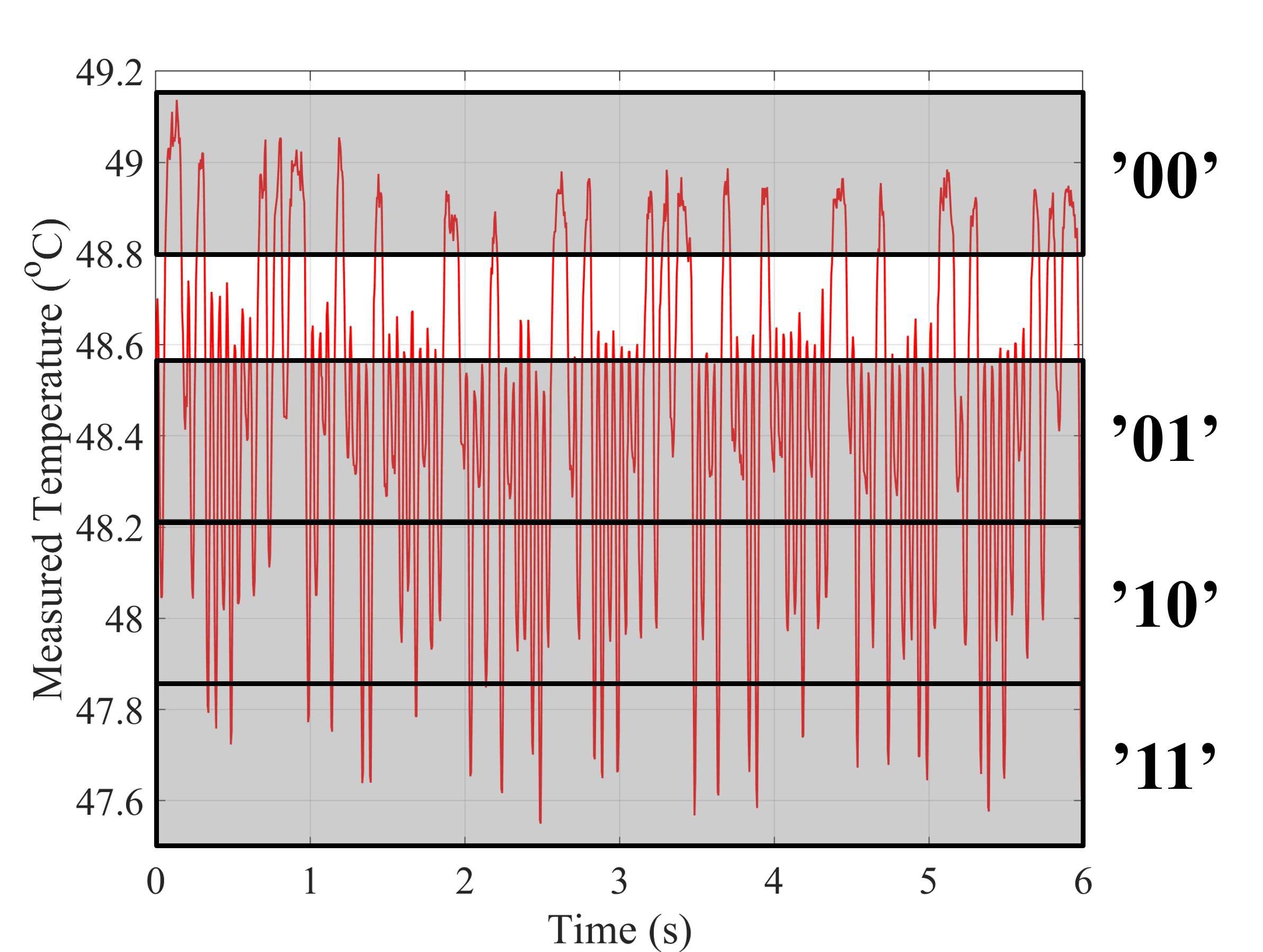}
  \caption{Resulting apparent temperature modulation from four level pulse amplitude modulation (PAM4).}
  \label{fig:pam}
\end{figure}

One way to improve the spectral efficiency of the communications channel, and relax the requirement on our receiver sampling rate is to move to higher order modulation schemes. We tested an implementation a four-level pulse amplitude modulation coding technique (PAM4), as shown in Fig. \ref{fig:pam}. From the electro-optic characterization described earlier, we further refined the duration of the applied voltage pulses to create four distinct levels of radiance that can be used to encode symbols. Fig. \ref{fig:pam} shows the recorded temperature value as a function of time, and delineates how these levels were decoded into PAM4 symbols.

To extrapolate our results to more realistic link ranges we built a communications link budget to compute achievable data rates as a function of transmitter and receiver parameters. Assuming far field operation, we can construct a communications link budget calculation by approximating our device to be a hemispherically isotropic point source radiating into $2\pi$ steradians. The radiation pattern can be modeled by the free space path loss (FSPL) equation, where the received power is given by $P_{R} = P_{T}A_{R} \exp{\left(-\alpha L\right)}/2\pi L^{2}$, where $P_{T}$ is the transmitted power, $A_{R}$ is the receiver aperture area, $\alpha$ is the attenuation coefficient for free-space path loss and $L$ is the range from transmitter to receiver. The FLIR camera acts as the optical receiver in the FSO link. We built the model on the assumption the noise is background limited and utilized the photostatistic curves shown in Fig. \ref{fig:Voltage_Signal}. Using this methodology, the standard deviation of the Gaussians will remain constant and the mean value will decrease by the FSPL equation. Estimating the probability of error using this model with our current setup, we can theoretically communicate below the forward error correction (FEC) threshold for Reed-Solomon Enhanced FEC ($<10^{-3}$) up to 28.7 meters, adding an overhead of $7\%$ to our required data rate. 

The achievable link range can be scaled dramatically by increasing the active area of the graphene modulators. For example, if we had a 1 m x 1 m sheet of graphene, the transmit area would be scaled by a factor of $10^{4}$ increasing the transmit power by by the same factor extending the error-free communications link reach to >2.5 km without improving any of our receiver parameters. Neither the encoding nor the decoding strategy for this communications channel has been optimized yet, and we envision important gains in data rate and link reach by addressing these topics.

While previous work generated engineered blackbody radiation from artificial sources to implement secure FSO communications over centimeter scale distances \cite{Lucyszyn}, we showed an implementation that currently scales to 28.7 meters, and can scale to kilometers in the future. Even works which used naturally occuring Planck radiation were limited to low-rate analog communications such as Morse code \cite{Ergoktas2020,guerra2020characterization} while our work demonstrates that this FSO communication system that can be scaled to higher rate digital communications. 

This technique shows promise for realizing broad spectrum ad-hoc and covert communications channels in the LWIR and the development of communications transmitters that can achieve small size, weight and power by harvesting naturally occuring energy for communications symbol generation.

%\section{Conclusion}
%We demonstrated a proof-of-concept FSO communications system harvesting Planck radiation to encode and transmit digital data. Future work can scale data rates and range by adding more graphene devices to the transmitter array and further investigating the ion dynamics of the devices. This technique shows promise for realizing broad spectrum ad-hoc and covert communications channels in the LWIR and the development of communications transmitters that can achieve small size, weight and power by harvesting naturally occuring energy for communications symbol generation.

\smallskip

\textbf{Funding} This research was supported by the Office of Naval Research (ONR) award no. N00014-22-1-2697 (H.W.), U.S. Department of Energy, Office of Basic Energy Sciences award DE-FG02-0746376 (Z.C.), and National Science Foundation (NSF) award no. CBET-2012845 (H.W.).  This work was also supported by a USC ISI exploratory research award. 

\textbf{Acknowledgments} The authors are grateful to Arunkumar Jagannathan for initial measurements on Planck modulator devices and experimental guidance and Megan S and Andrew K for their help fabricating graphene. 

\smallskip

\textbf{Disclosures} The authors declare no conflicts of interest.

\textbf{Data availability} Data underlying the results presented in this paper are not publicly available at this time but may be obtained from the authors upon reasonable request.
\bibliographystyle{unsrt}
\bibliography{lwircomms}

\end{document}